\documentclass[a4paper]{jpconf}
\usepackage{graphicx}
\usepackage{hyperref}
\begin{document}
\title{Upsilon Production in Pb-Pb and p-Pb Collisions at Forward Rapidity with ALICE at the LHC}

\author{Palash Khan for the ALICE Collaboration}

\address{Saha Institute of Nuclear Physics, Kolkata, India}

\ead{palash.khan@cern.ch}

\begin{abstract}
The ALICE apparatus at the LHC was designed and built to perform dedicated studies of the Quark-Gluon Plasma (QGP), a strongly interacting phase of QCD matter, expected to be created in heavy-ion collisions, where quarks and gluons are deconfined. In such collisions heavy flavours are produced at the very early stage of the interaction by the initial hard scattering processes and hence can be used to characterize the hot and dense medium. In particular the sequential suppression of quarkonia (charmonia and bottomonia) was proposed as a thermometer of the deconfined medium. The inclusive $\Upsilon(1S)$ production has been measured down to zero transverse momentum in its dimuon decay channel at forward rapidity $(2.5 < y_{\rm _{lab}} < 4.0)$ using the Muon Spectrometer. Here results on the $\Upsilon(1S)$ nuclear modification factor $(R_{\rm AA})$ in Pb-Pb collisions at $\sqrt{s_{\rm NN}}$ = 2.76 TeV are discussed and compared to the measurement at mid-rapidity by the CMS Collaboration and to theoretical predictions. Also recent results on $R_{\rm pPb}$ and forward-to-backward yield ratio $(R_{\rm FB})$ in p-Pb collisions at $\sqrt{s_{\rm NN}}$ = 5.02 TeV are discussed.

\end{abstract}

\section{Introduction}
\hspace{5 mm} At very high temperature and energy density the hadronic matter can turn into a state of deconfined quarks and gluons known as “Quark-Gluon Plasma” (QGP)~\cite{E. V. Shuryak}. This state of matter can be created by colliding two large nuclei at ultra-relativistic energies. In such collisions, heavy flavours, specially charm and bottom, are produced at the very early stage by the hard scattering and hence can be used to characterize the hot and dense medium. It is predicted that the production of quarkonia (charmonia and bottomonia) in nucleus-nucleus collisions will be suppressed relative to that in proton-proton due to the Color-Debye Screening mechanism~\cite{T. Matsui et al}. In particular, a sequential suppression of quarkonia, depending on their binding energy has been proposed as a thermometer of the deconfined medium~\cite{S. Digal et al}. 

The suppression of quarkonia can be quantified by measuring the nuclear modification factor $R_{\rm AA}$, which is the ratio of yield in nucleus-nucleus collisions to that in proton-proton collisions scaled by the number of binary collisions.

The suppression of quarkonium states depends on several factors, as the feed-down contributions from higher-mass resonances into the observed quarkonium yield, the b-hadrons decay into charmonium and the quarkonium regeneration due to the q$\bar{\rm q}$ recombination (expected to be important for c$\bar{\rm c}$ pairs at LHC energies)~\cite{A. Andronic et al}. Other important competing mechanisms are the Cold Nuclear Matter (CNM) effects, such as nuclear shadowing or gluon saturation, which can break the binary scaling even in absence of the QGP. Hence, for the interpretation of results in heavy-ion collisions, data from pA are crucial since they allow to disentangle the CNM effects from those related to the formation of a hot medium. With this goal in mind, the ALICE~\cite{K. Aamodt et al} collaboration has measured the $\Upsilon(1S)$ production in Pb-Pb collisions at $\sqrt{s_{\rm NN}}$ = 2.76~TeV and p-Pb collisions at $\sqrt{s_{\rm NN}}$ = 5.02~TeV. For p-Pb, data were taken with two beam configurations corresponding to the proton going towards or opposite to the direction of the muon spectrometer (labelled p-Pb and Pb-p in the following).

\section{Results}
\begin{figure}[h]
\begin{minipage}{18pc}
\includegraphics[width=18pc]{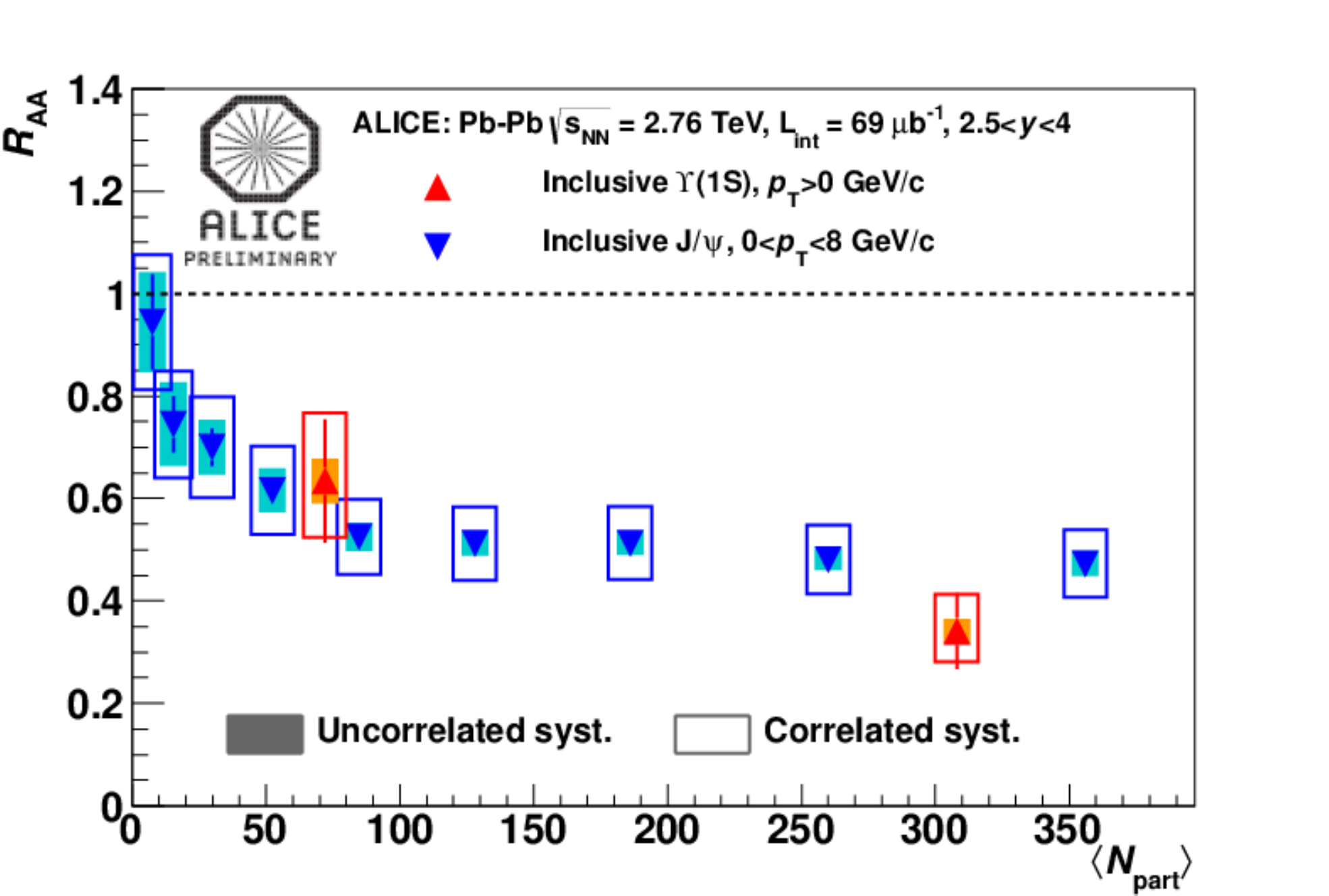}
\end{minipage}\hspace{2pc}
\begin{minipage}{18pc}
\includegraphics[width=18pc]{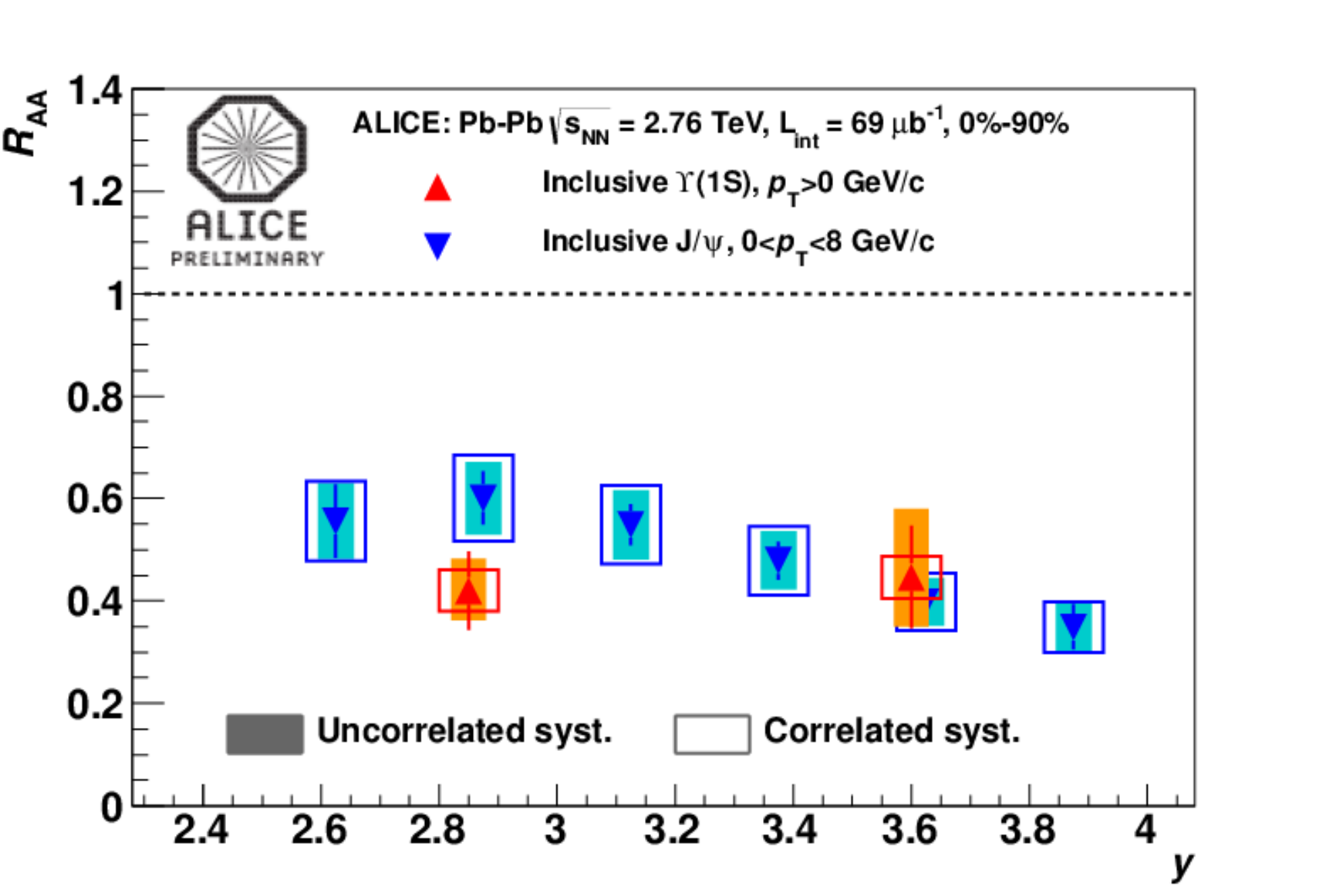}
\end{minipage} 
\caption{\label{label}$R_{\rm AA}$ as a function of $\langle N_{\rm part} \rangle$ and $y$ for J/$\psi$ and $\Upsilon(1S)$ as measured by ALICE.}
\label{fig:RaaVsNpartAndRapidityJpsiAndUpsilon}
\end{figure}

\hspace{5 mm} A data sample corresponding to an integrated luminosity of $L_{\rm int}$ = 69.2 $\mu {\rm b^{-1}}$ (for Pb-Pb), 5.3 ${\rm nb^{-1}}$ (for p-Pb) and 6.1 ${\rm nb^{-1}}$ (for Pb-p) have been collected. In Pb-Pb collisions, the measurement of the nuclear modification factor of the inclusive $\Upsilon(1S)$ production has been performed in the $2.5 < y_{\rm cms} < 4$ rapidity range, down to $p_{\rm T}$ = 0 and in the $0\%-90\%$ centrality class at $\sqrt{s_{\rm NN}}$ = 2.76~TeV. The integrated $R_{\rm AA}$ value is 0.439$\pm$0.065(stat.)$\pm$0.028(uncorrelated syst.)$^{+0.092}_{-0.076}$(correlated syst.) and implies a significant suppression of inclusive $\Upsilon(1S)$. Two centrality ranges were studied and a stronger suppression in more central collisions has been observed (left plot of Fig.~\ref{fig:RaaVsNpartAndRapidityJpsiAndUpsilon}). In addition, two rapidity ranges were studied and no significant variation of the suppression with rapidity was observed (right plot of Fig.~\ref{fig:RaaVsNpartAndRapidityJpsiAndUpsilon}). The $\Upsilon(1S)$ $R_{\rm AA}$ as a function of both rapidity and centrality was found to be comparable with that observed for the J/$\psi$ measured by ALICE in the same kinematic range (left and right plot of Fig.~\ref{fig:RaaVsNpartAndRapidityJpsiAndUpsilon}). The interpretation of this observation is not straightforward due to the different amount of feed down from higher mass states and also due to the presence of a recombination contribution for the J/$\psi$.

\begin{figure}[h]
\begin{minipage}{18pc}
\includegraphics[width=18pc]{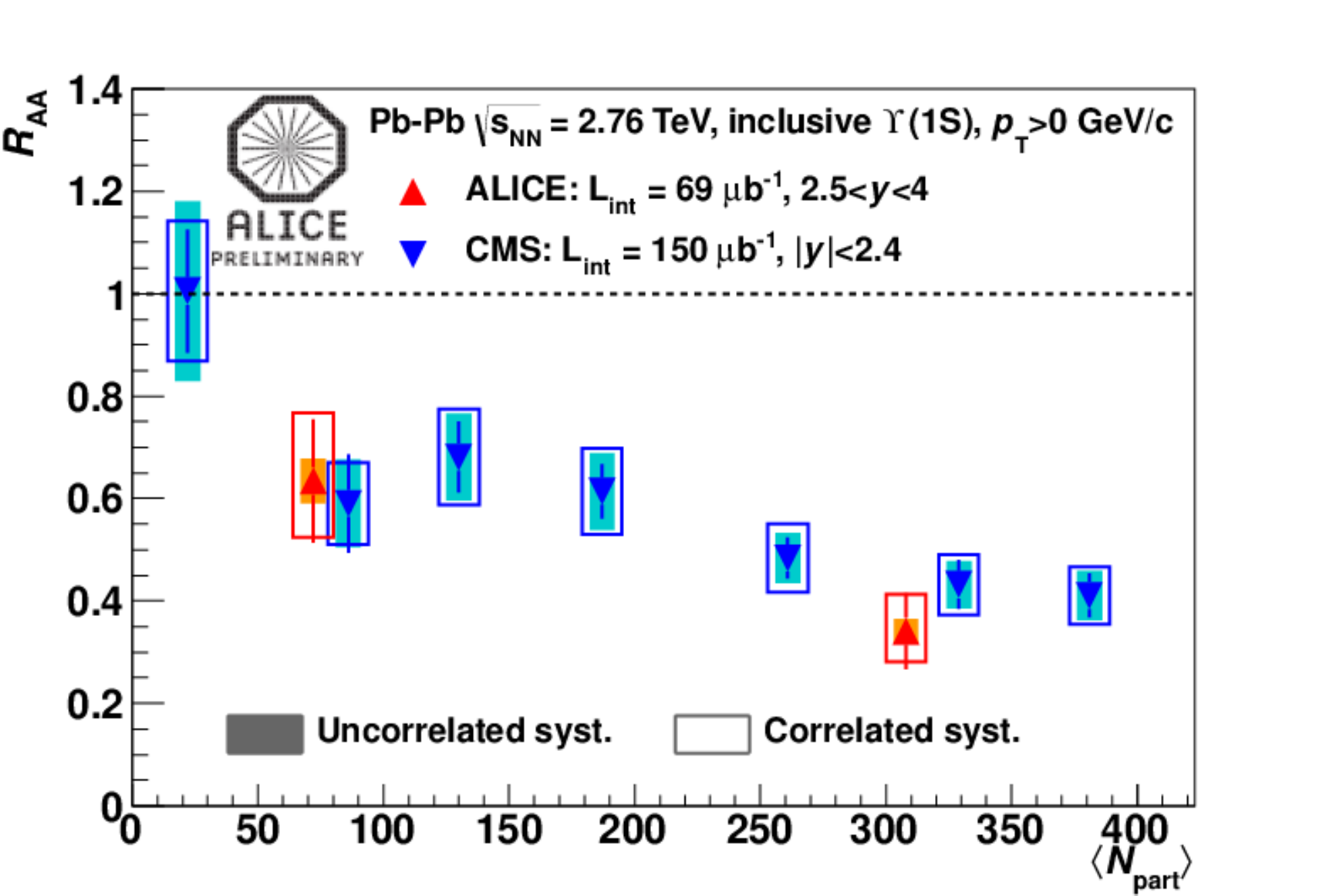}
\end{minipage}\hspace{2pc}
\begin{minipage}{18pc}
\includegraphics[width=18pc]{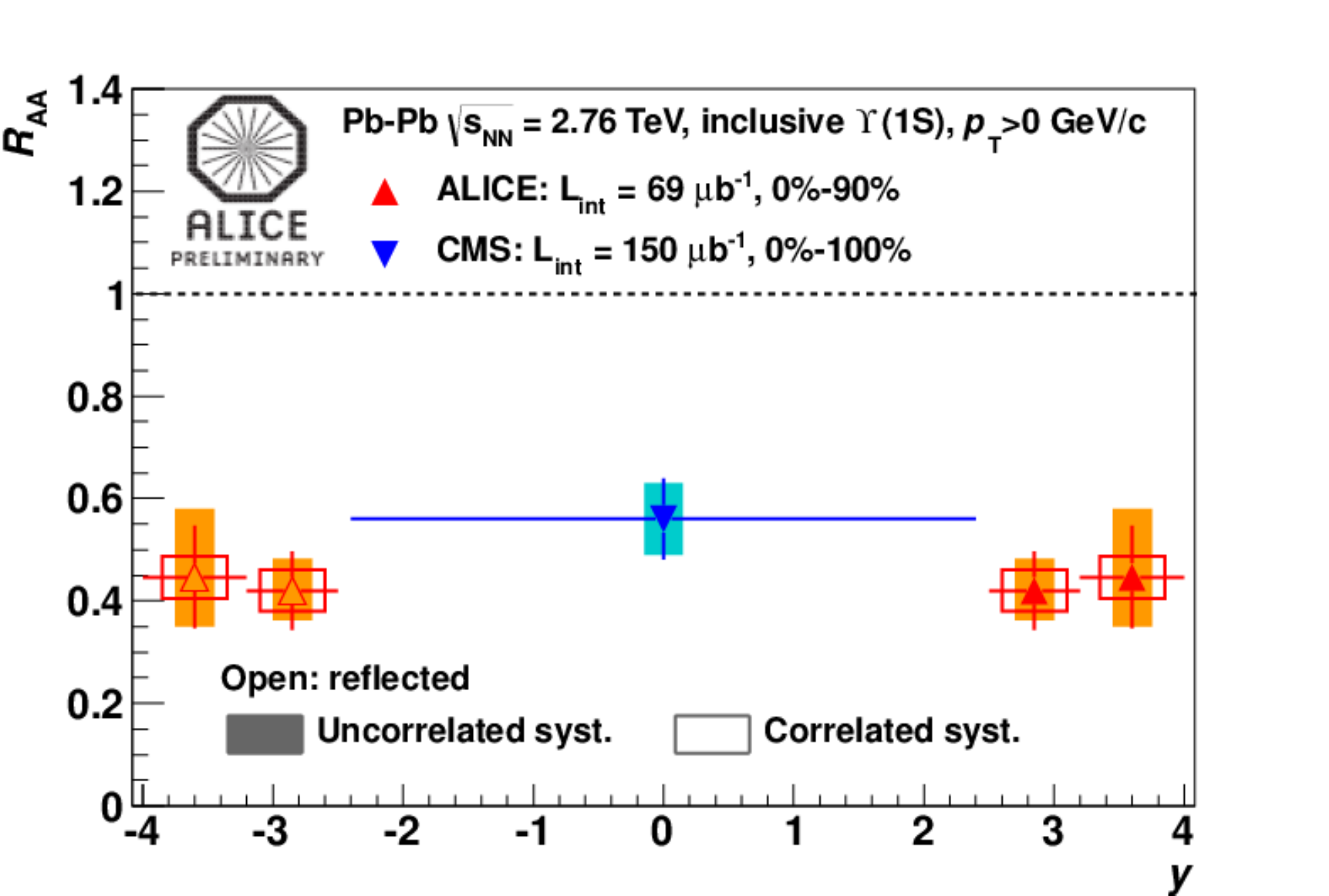}
\end{minipage} 
\caption{\label{label}$R_{\rm AA}$ as a function of $\langle N_{\rm part} \rangle$ and $y$ for $\Upsilon(1S)$ as measured by ALICE and CMS.}
\label{fig:RaaVsNpartAndRapidityALICEandCMS}
\end{figure}

The $\Upsilon(1S)$ $R_{\rm AA}$ is compared to CMS results for the kinematic range $|y| < 2.4$~\cite{S. Chatrchyan et al} (left plot of Fig.~\ref{fig:RaaVsNpartAndRapidityALICEandCMS}). A similar suppression as a function of centrality has been found by both ALICE and CMS. Small variation with rapidity of the inclusive $\Upsilon(1S)$ suppression has been observed over the present range accessed by the two experiments (right plot of Fig.~\ref{fig:RaaVsNpartAndRapidityALICEandCMS}).

\begin{figure}[h]
\begin{minipage}{18pc}
\includegraphics[width=18pc]{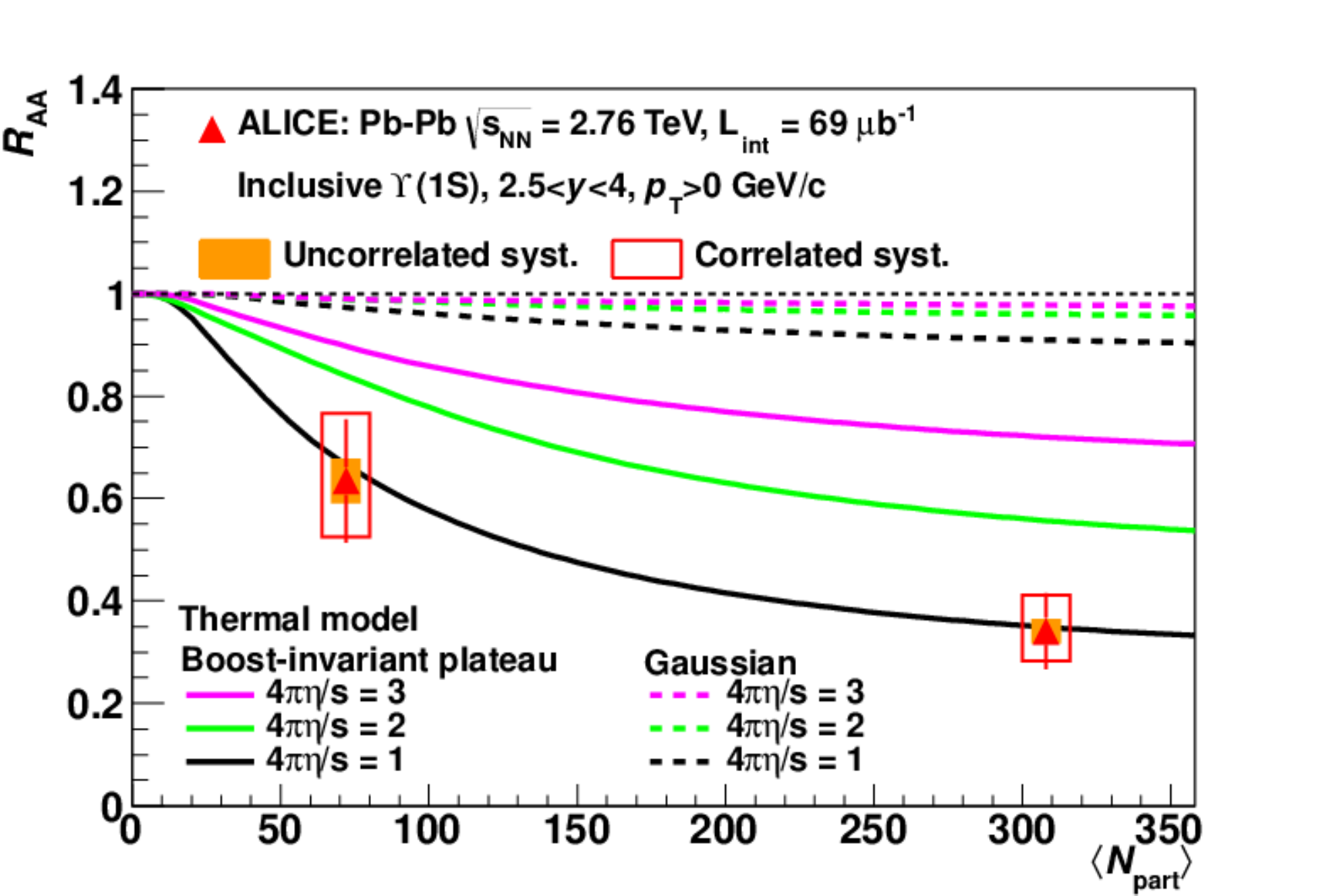}
\end{minipage}\hspace{2pc}
\begin{minipage}{18pc}
\includegraphics[width=18pc]{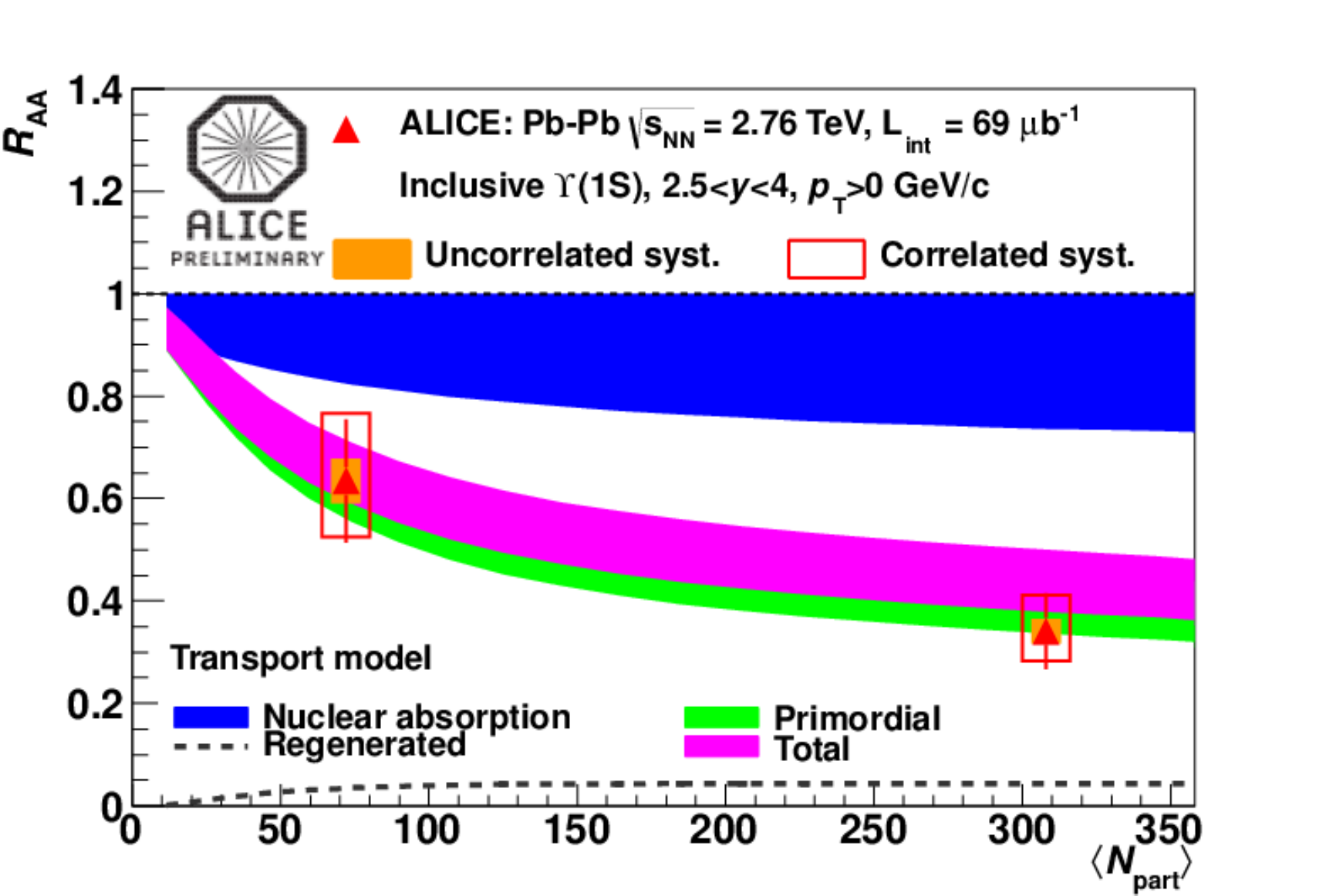}
\end{minipage} 
\caption{\label{label}ALICE $R_{\rm AA}$ as a function of $\langle N_{\rm part} \rangle$ for $\Upsilon(1S)$ compared with models from Strickland~\cite{M. Strickland} and Emerick et al.~\cite{A. Emerick et al}.}
\label{fig:RaaVsNpartStricklandAndEmerick}
\end{figure}

Our results are also compared with models. The Anisotropic Hydro Model~\cite{M. Strickland} includes feed-down of $\Upsilon(1S)$ by higher mass states, but does not include recombination effects nor cold nuclear matter effects. Data is described with the hypothesis of a boost invariant plateau temperature profile with minimum shear viscosity at forward-rapidity (left plot of Fig.~\ref{fig:RaaVsNpartStricklandAndEmerick}). The Transport Model~\cite{A. Emerick et al} includes small b$\bar{\rm b}$ regeneration, feed-down from higher mass (around 50 \%) and CNM effects by an overall absorption cross-section of 0-2 mb. The model is in fair agreement with data within uncertainties (right plot of Fig.~\ref{fig:RaaVsNpartStricklandAndEmerick}).

\begin{figure}[h]
\begin{minipage}{18pc}
\includegraphics[width=18pc]{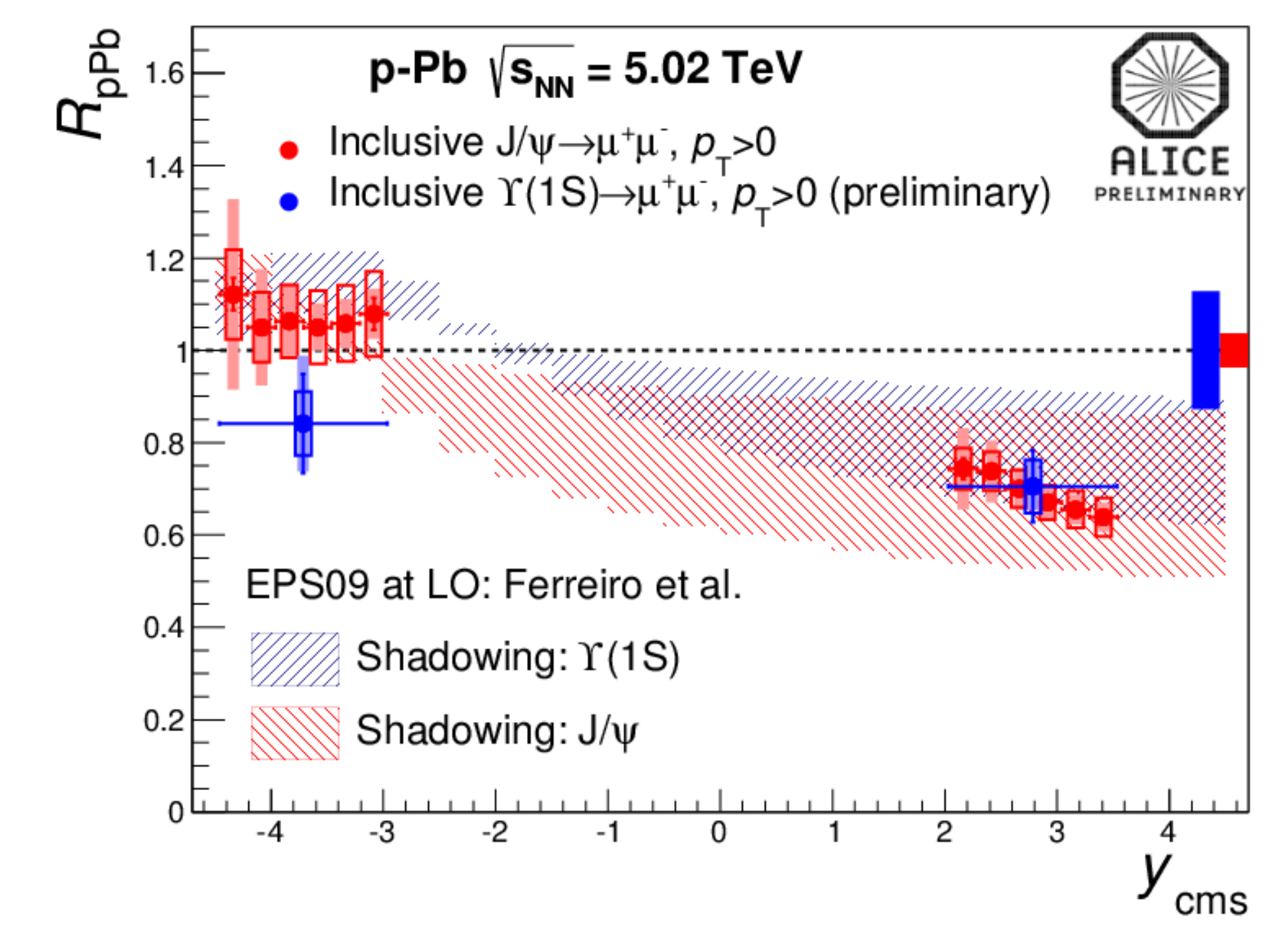}
\end{minipage}\hspace{2pc}
\begin{minipage}{18pc}
\includegraphics[width=18pc]{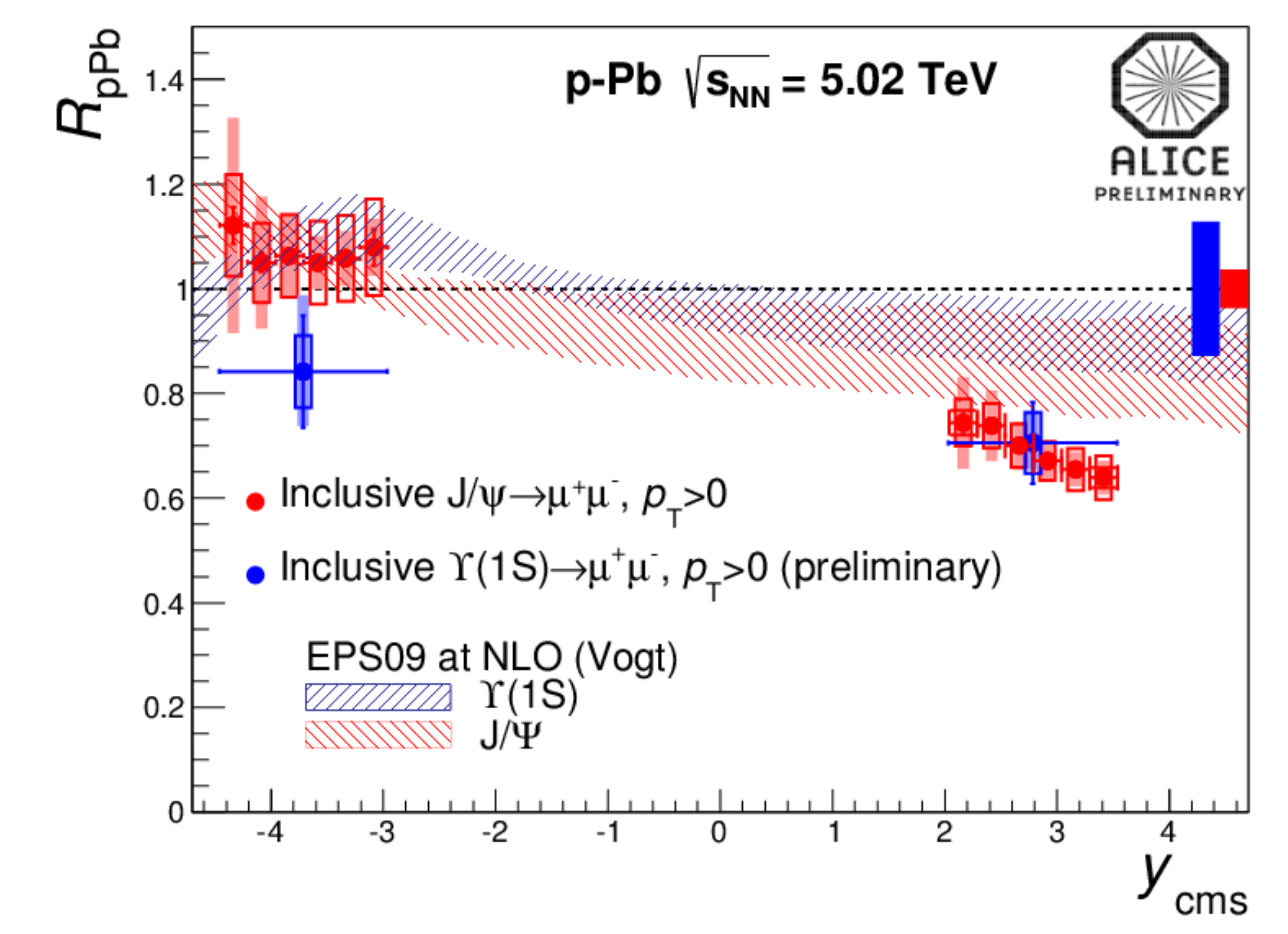}
\end{minipage} 
\caption{\label{label}Comparison of the inclusive $\Upsilon(1S)$ $R_{\rm pPb}$ with inclusive J/$\psi$ data and models from Ferreiro et al.~\cite{E. G. Ferreiro et al} and Vogt et al.~\cite{J. L. Albacete et al}.}
\label{fig:RppbVsRapidityFerreiroAndVogt}
\end{figure}

In case of p-Pb and Pb-p the inclusive $\Upsilon(1S)$ $R_{\rm pPb}$ was measured at forward $(2.04 < y_{cms} < 3.54 )$ and backward rapidity $(-4.46 < y_{cms} < -2.96)$. A stronger suppression is observed for inclusive $\Upsilon(1S)$ at the forward rapidity than backward rapidity (Fig.~\ref{fig:RppbVsRapidityFerreiroAndVogt}). The results are compared to that of the J/$\psi$ in the same kinematic range. The observed suppression for the $\Upsilon(1S)$ is comparable with the J/$\psi$ one at forward rapidity, while at backward rapidity the suppression of $\Upsilon(1S)$ is stronger.

The experimental results are compared with two models. The EPS09 Shadowing calculation at LO~\cite{E. G. Ferreiro et al} and NLO~\cite{J. L. Albacete et al} agrees reasonably well with the data at forward rapidity but tend to underestimate the $\Upsilon(1S)$ suppression at backward rapidity (left and right plot of Fig.~\ref{fig:RppbVsRapidityFerreiroAndVogt}). Within the present uncertainties, neither of the calculations are favoured by the data.

\begin{figure}[h]
\begin{minipage}{18pc}
\includegraphics[width=18pc]{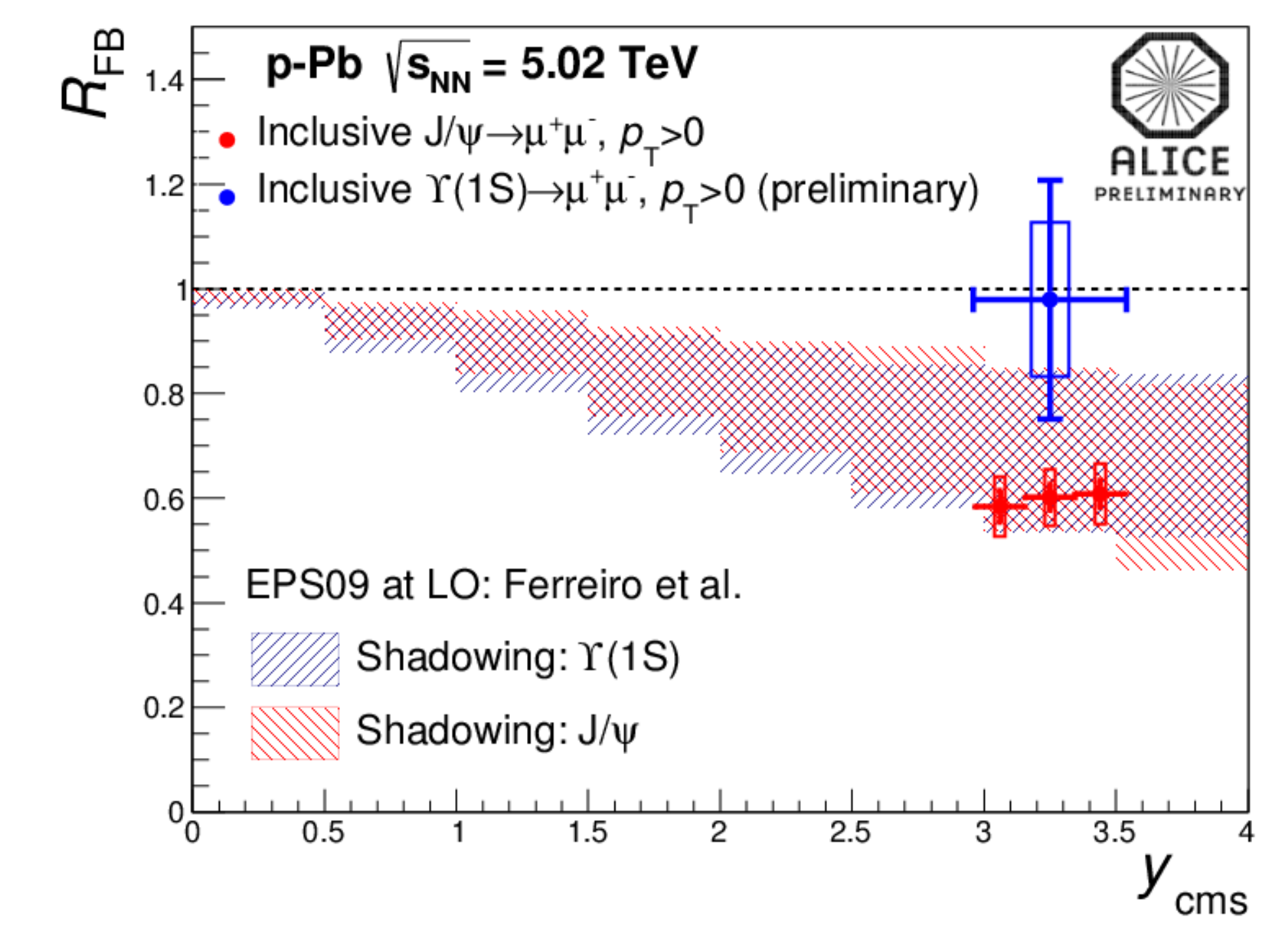}
\end{minipage}\hspace{2pc}
\begin{minipage}{18pc}
\includegraphics[width=18pc]{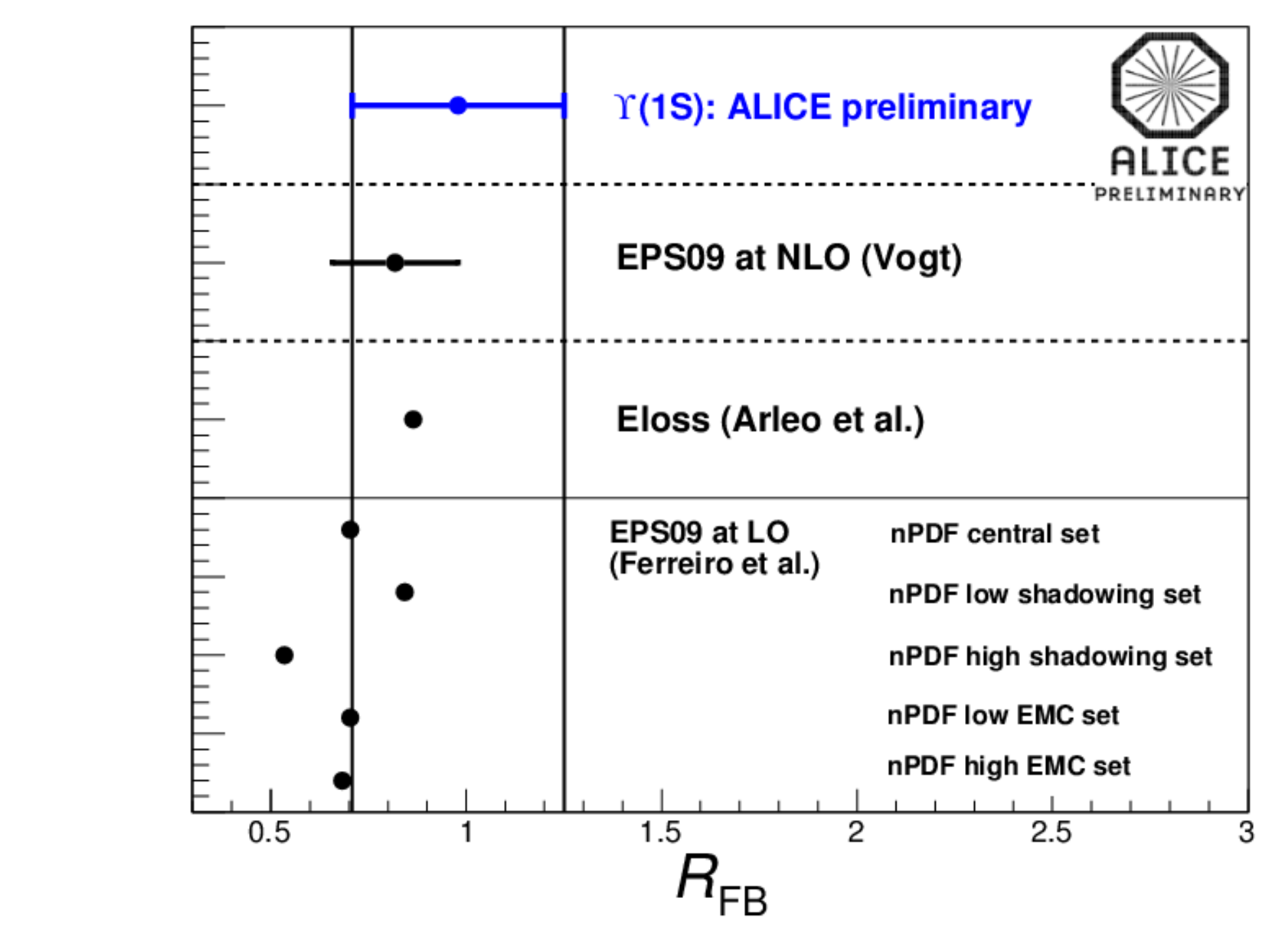}
\end{minipage} 
\caption{\label{label}Comparison of the inclusive $\Upsilon(1S)$ $R_{\rm FB}$ with various models~\cite{J. L. Albacete et al},~\cite{F. Arleo et al},~\cite{E. G. Ferreiro et al}.}
\label{fig:RfbVariousModels}
\end{figure}

The ratio $R_{\rm FB}$ between forward and backward $\Upsilon(1S)$ production was measured in the common rapidity region $(2.96 < |y_{cms}| < 3.54)$ of p-Pb and Pb-p. It is found to be larger than that of the J/$\psi$ measured in the same kinematic window (left plot of Fig.~\ref{fig:RfbVariousModels}). The results are compared with an EPS09 shadowing calculation at both LO~\cite{E. G. Ferreiro et al} and NLO~\cite{J. L. Albacete et al} and with the Coherent Parton Energy Loss model~\cite{F. Arleo et al} (right plot of Fig.~\ref{fig:RfbVariousModels}). The agreement is reasonably good, except for the LO case with larger shadowing.

\section{Conclusions}
\hspace{5 mm}The nuclear modification factor for inclusive $\Upsilon(1S)$ has been measured at forward rapidity $2.5 < y < 4.0$ down to zero $p_{\rm T}$. In Pb-Pb collisions, the suppression of inclusive $\Upsilon(1S)$ is stronger in central than semi-peripheral collisions. The $\Upsilon(1S)$ suppression pattern is comparable with forward-rapidity J/$\psi$ result from ALICE within uncertainties. No strong variation with rapidity has been observed within the large range probed by ALICE and CMS. 

\hspace{5 mm}In p-Pb data, we observe a small suppression of $\Upsilon(1S)$, which tends to increase from backward to forward rapidity. At forward rapidity the J/$\psi$ suppression is comparable with the one of $\Upsilon(1S)$ within uncertainties. The J/$\psi$  $R_{\rm FB}$ is significantly lower than the one measured for $\Upsilon(1S)$.

\end{document}